\documentclass[aip,jcp,floatfix,reprint]{revtex4-1}
\usepackage{amsmath}
\usepackage{graphicx}
\usepackage{amsfonts}
\usepackage{color}
\usepackage{placeins}

\newcommand{\rr}{{\bf r}}
\newcommand{\RR}{{\bf R}}

\newcommand{\PP}{\mathcal{P}}
\newcommand{\eeps}{\boldsymbol\epsilon}
\newcommand{\rrho}{\boldsymbol\rho}

\begin{document}

\title{Optimization of an exchange-correlation density functional for water}

\author{ Michelle Fritz }
\affiliation{ Departamento de F\'{\i}sica de la Materia Condensada,
              Universidad Aut\'{o}noma de Madrid,
              E-28049 Madrid, Spain }

\author{ Marivi Fern\'andez-Serra }
\affiliation{ Department of Physics and Astronomy,
              Stony Brook University, 
              Stony Brook, New York 11794-3800, USA }
\affiliation{ Institute for Advanced Computational Science, 
              Stony Brook University,
              Stony Brook, New York 11794-3800, USA}

\author{ Jos\'e M. Soler }
\affiliation{ Departamento e Instituto de F\'{\i}sica de la Materia Condensada,
              Universidad Aut\'{o}noma de Madrid,
              E-28049 Madrid, Spain }

\date{\today}

\begin{abstract}
   
   We describe a method, that we call data projection onto parameter space 
(DPPS), to optimize an energy functional of the electron density, so that 
it reproduces a dataset of experimental magnitudes. 
   Our scheme, based on Bayes theorem, constrains
the optimized functional not to depart unphysically from existing 
{\it ab initio} functionals.
   The resulting functional maximizes the probability of being the ``correct"
parametrization of a given functional form, in the sense of Bayes theory.
   The application of DPPS to water sheds new light on why density functional
theory has performed rather poorly for liquid water, on what improvements
are needed, and on the intrinsic limitations of the generalized gradient
approximation to electron exchange and correlation.
   Finally, we present tests of our water-optimized functional,
that we call vdW-DF-w, showing that it performs very well for 
a variety of condensed water systems.

\end{abstract}

\maketitle

\section{Introduction}


 Liquid water is arguably the most important substance for life, as well
as for an immense number of problems of huge scientific and technological
importance\cite{Hummer2014,Walter2010}.
   At the same time, despite its molecular simplicity, it is a liquid of astonishing 
complexity, with tens of thermodynamic anomalies.
   Ultimately, this complexity steams from the coexistence of covalent,
electrostatic, and dispersion interactions, that have very different magnitudes
but also subtle and critical interrelations\cite{Wang2011,Pamuk2012,Corsetti2013b}.
   Thus, it not surprising that electron density functional theory (DFT),
a universal and completely non-empirical method, has had a 
particularly hard time in describing all these interactions with the
required accuracy\cite{Wang2011}.
   In fact, at present, DFT simulations cannot match the success of
empirical force fields in simulating a wide range of effects and
anomalies\cite{Vega2009,Vega2011}.
   However, empirical methods are not necessarily reliable outside
the range where they have been fitted, like in deeply undercooled
water, a sate in which it has been predicted to have a liquid-liquid
transition that could explain many of its anomalies at higher temperatures\cite{Poole1992,Xu2005,Abascal2010}.
   Therefore, its accurate description by DFT remains a very important 
challenge to understand the intricate structure and properties of
liquid water.
   In this work we aim at uncovering in detail the deficiencies of
present functionals and at developing an optimized functional 
within the generalized gradient approximation (GGA).


   The parameterization and optimization of complex models is a
pervasive problem in many areas, and in particular in the development
of interatomic potentials and functionals for molecular dynamics.
   Generally, it requires to choose largely arbitrary functional forms 
that depend on many parameters, followed by a lengthy and difficult
trial and error optimization\cite{Medders2014}.
   The balance between the number of parameters and the size of the
fitted data sets involves difficult and subjective decisions that are
nevertheless critical to the results.
   Here we describe a general and powerful optimization scheme, 
data projection onto parameter space (DPPS), and its application to
the optimization of an exchange-correlation (xc) energy functional of the
electron density.

\section{Data projection onto parameter space}


   Before addressing functional optimization, it is useful to consider
the method from a broader perspective.
   DPPS tries to find the optimal parameters for a complex model that
``predicts'' (calculates) a scalar function $E$ (say the potential energy) 
that depends on a large number of variables $\RR_i$ (say the atomic positions) 
and parameters.
   To be specific, imagine that we want to fit a pairwise interatomic potential 
$V(R)$ (assuming a single chemical species).
   The first step would be to choose a number of radial interpolation
mesh points $R_\alpha$ for the interatomic distance.
   The parameters of the model would then be the values 
$\epsilon_\alpha \equiv V(R_\alpha)$, from which we would interpolate 
$V(R) = \sum_\alpha \epsilon_\alpha p_\alpha(R)$, where $p_\alpha(R)$ are
a suitable set of interpolation basis functions.
   They are unambiguosly determined by the interpolation scheme (e.g.
cubic splines~\cite{NumericalRecipes,splines}) and by the interpolation points: 
thus $p_\alpha(R)$ is the result of interpolating a function $f(R)$ with 
$f(R_\beta)=\delta_{\alpha\beta}$.
   Thus, $p_\alpha(R_\beta)=\delta_{\alpha \beta}$ and, for a sufficiently 
fine mesh, $p_\alpha(R) \simeq \delta(R-R_\alpha$).

   The next step is to set up a large data set of known system geometries
$\RR_i^n$ (where $i$ denotes atoms and $n$ denotes systems or geometries) 
and associated energies $E_{ref}^n$, obtained from experiments 
or from higher level calculations.
    The intended prediction of the model would then be
\begin{equation}
 E_{ref}^n = E^n \equiv \sum_{i<j} V(R^n_{ij}) 
    = \sum_\alpha \epsilon_\alpha \sum_{i<j} p_\alpha(R^n_{ij})
    \equiv \eeps \bullet \rrho^n
 \label{Eofv}
\end{equation}
where $\eeps \equiv \{\epsilon_\alpha\}$ and $\rrho^n \equiv \{\rho^n_\alpha\}$,
are vectors in ``parameter space", with
$\rho^n_\alpha \equiv \sum_{i<j} p_\alpha(R^n_{ij})$ 
being proportional to the radial distribution function 
(i.e. the radial interatomic density) at the mesh distances.

   Eq.~(\ref{Eofv}) sets the projections $E^n_{ref}$ of the vector $\eeps$ 
of unknown parameters onto the vectors $\rrho^n$ of known data.
   Depending on the relative numbers of parameters and data,
$N_{dat}$ and $N_{par}$,
the later will form a subspace of the former or they will overdetermine them.
   If $N_{dat}>N_{par}$, the system (\ref{Eofv}) can be solved in the least
squares minimization sense, the solution being
$\epsilon_\alpha = \sum_\beta S^{-1}_{\alpha \beta} u_\beta$, where
$S_{\alpha \beta} \equiv \sum_n \rho_\alpha^n \rho^n_\beta$ and 
$u_\beta \equiv \sum_n E_{ref}^n \rho^n_\beta$.

   In practice, even if $N_{dat}>>N_{par}$, all the data may nearly lie in a 
subspace of the parameter space.
   As a result, certain combinations of parameters may be poorly determined,
and very large parameter values will result if we require an exact fit of the
data energies.
   This situation will be apparent by a nearly singular matrix 
$S_{\alpha \beta}$, the standard cure being to invert it by singular value
decomposition, discarding the subspace with small eigenvalues.
   Another standard alternative
is to add a regularization penalty to large parameter changes, minimizing
\begin{equation}
Z = \sum_{n=1}^{N_{dat}} \left( \frac{E^n-E_{ref}^n}{\Delta E_n} \right)^2 + 
    \sum_{\alpha=1}^{N_{par}} 
    \left( \frac{\epsilon_\alpha-\epsilon^0_\alpha}
                {\Delta \epsilon_\alpha} \right)^2 = \min
\label{Z}
\end{equation}
where $\Delta E_n$ are error estimates of $E_{ref}^n$, and $\epsilon^0_\alpha$,
$\Delta \epsilon_\alpha$ are initial estimates of the parameter values and their
uncertainties.
   In practice, using $\epsilon^0_\alpha=0$ and constant values for $\Delta E_n$
and $\Delta \epsilon_\alpha$ is essentially equivalent to singular value 
decomposition, with an eigenvalue cutoff $\sim \Delta E / \Delta \epsilon$.
   However, Eq.~(\ref{Z}) provides a smoother 
and more natural transition to the Bayesian approach described below.

   If no initial estimate $\epsilon^0_\alpha$ is available, one can instead 
impose smoothness of $V(R)$, with a penalty to the first or second derivatives.
   Such a penalty can be rationalized, as in the Gaussian regression 
method~\cite{Rasmussen-Williams2006}, as prior Bayesian information on 
the function smoothness.

\section{Functional optimization}

   Consider now the optimization of an xc energy functional 
of the electron density $\rho(\rr)$, 
in the generalized gradient approximation (GGA):
$E_{xc}[\rho(\rr)] = \int 
   \rho(\rr)~\epsilon_{xc}\left(k_F(\rr),k_G(\rr)\right)~d^3\rr$,
where $k_F=(3\pi^2 \rho)^{1/3}$ and $k_G=|\nabla\rho|/\rho$.
   We use these two wavevectors as functional variables, rather than the
conventional $\rho$ and $s=k_G/2k_F$, because they have the same dimension
and similar magnitude (see below) and therefore they provide a more 
``isotropic'' parameter space.
   To parametrize $\epsilon_{xc}(k_F,k_G)$, 
we choose interpolation points $k_{F\alpha}$ and $k_{G\beta}$ and we expand 
$\epsilon_{xc}(k_F,k_G) = \sum_{\alpha\beta}~\epsilon_{\alpha\beta} 
~p_\alpha(k_F)~p_\beta(k_G)$, so that $\epsilon_{\alpha\beta}$ are our 
functional parameters. 
   Assuming again that we have a dataset of system geometries $\RR_i^n$ and 
corresponding total energies $E^n_{ref}$, we start with initial values
$\epsilon^0_{\alpha\beta}=\epsilon^0_{xc}(k_{F \alpha},k_{G \beta})$,
where $\epsilon^0_{xc}(k_{F},k_{G})$ is a 
reference functional~\cite{Perdew-Burke-Ernzerhof1996}.
   We find the self-consistent electron densities $\rho_n(\rr)$ and the initial
total and xc energies, $E^{0n}_{tot}$ and $E^{0n}_{xc}$, for each system $n$.
   Then, to first order in $\delta \epsilon_{\alpha\beta}$,
\begin{eqnarray}
\delta E_{tot} 
   &\simeq& \sum_{\alpha\beta} \left( \frac{\partial 
      E_{tot}}{\partial \epsilon_{\alpha\beta}} + 
      \int \frac{\delta E_{tot}}{\delta \rho(\rr)} 
        \frac{\partial \rho(\rr)}{\partial \epsilon_{\alpha\beta}} 
       d^3\rr \right) \delta \epsilon_{\alpha\beta} \nonumber \\
   &=& \sum_{\alpha\beta} 
        \frac{\partial E_{xc}}{\partial \epsilon_{\alpha\beta}}
        \delta \epsilon_{\alpha\beta}
    \simeq \delta E_{xc}
\label{DEtot}
\end{eqnarray}
where we have used that $\delta E_{tot} / \delta \rho(\rr)=0$ and
$\partial E_{tot} / \partial \epsilon_{\alpha\beta} =
 \partial E_{xc} / \partial \epsilon_{\alpha\beta}$.
   Thus, we impose that
$E^n_{ref}-E^{0n}_{tot} = E^n_{xc}-E^{0n}_{xc}$, or $E^n_{refxc}=E^n_{xc}$,
where
\begin{equation}
E^n_{refxc} \equiv E^n_{ref} - E^{0n}_{tot} + E^{0n}_{xc}
\end{equation}
and
\begin{equation}
E^n_{xc} \equiv \sum_{\alpha,\beta} \epsilon_{\alpha\beta}
       \int d^3\rr ~\rho_n(\rr) p_\alpha(k_F(\rr)) p_\beta(k_G(\rr))
   = \eeps \bullet \rrho^n
\label{Enxc}
\end{equation}
where $\eeps=\{\epsilon_{\alpha\beta}\}$, $\rrho^n=\{\rho^n_{\alpha\beta}\}$,
and $\rho^n_{\alpha\beta}$ are the integrals in (\ref{Enxc}).
   They are the density of electrons in parameter space, and
they are closely related to the functions $g_1(r_s)$ and $g_3(s)$ 
of Zupan {\it et al}~\cite{Zupan-Burke-Ernzerhof-Perdew1997}.

   Although now the $\epsilon_{\alpha\beta}$'s span a 2-D interpolation grid
(or a higher-dimensional mesh in case of a more complicated functional form, 
like meta-GGAs), it is convenient for notational simplicity to order them as 
a vector, using a single index index $\alpha$ for each pair of values 
$(k_F,k_G)_\alpha$,
   Thus, it is clear that Eqs.~(\ref{Eofv}) and (\ref{Enxc}) are 
entirely equivalent.
   Then, we can solve (\ref{Z}) to find new functional 
parameters $\epsilon_{\alpha}$ (i.e. a new xc functional)
and iterate to obtain selfconsistency between 
$\epsilon_{\alpha}$, $E^n_{xc}$, and $\rho(\rr)$.

   For exchange-only functionals, a simple scaling law requires
the GGA exchange energy density to have the form
$\epsilon_x^{GGA}(k_F,k_G) = \epsilon_x^{LDA}(k_F) F_x(s)$,
where $\epsilon^{LDA}_x(k_F)=-3k_F/4\pi$ is the energy density in
the local density approximation (LDA) and $s \equiv k_G/2k_F$ is
a reduced adimensional gradient.
   Thus, like in the pair-potential example, in this case we must optimize 
the single-variable function $F_x(s)$, usually called ``enhancement factor''.

   In general, known experimental data refer to energy differences rather
than total energies.
   Thus we may know accurately reaction and atomization energies,
or energy differences between different solid phases.
   To use these data efficiently, it suffices to substitute $E_n$ and
$\epsilon_n$ by $\Delta E_n$ and $\Delta \epsilon_n$ in the objective
function $Z$ and in all the equations above, where $\Delta$ refers to
the difference between the two systems or geometries.
   This procedure also allows to use a large variety of structural and 
thermodynamic data.
   Thus, to impose known equilibrium geometries, we consider two 
geometries with one of the atoms displaced by $\pm \Delta R$.
   Since the force must be zero, $\Delta \epsilon_n=0$.
   The same can be done to impose zero pressure and stress at the
known equilibrium geometry of a solid.

   Equally, known vibration frequencies $\omega_q$ can be imposed using
frozen-phonon displacements. 
   Alternatively, a larger set of random geometries and corresponding
energies (possibly taken from a molecular dynamics simulation)
can be used to optimize both the equilibrium geometry and the
deformation energies, beyond the harmonic approximation.

   Electronic data can be imposed through 
constrained DFT~\cite{Kaduk-Kowalczyk-VanVoorhis2012}.
   For dipole moments and polarizabilities, the Lagrange multiplier associated
to the electronic constraint is simply an external electric field.
   Thus, these constraints can be set similarly to those of the equilibrium
geometry and deformation frequencies.

\section{Bayesian optimization}

   Although the simple penalty term of Eq.~(\ref{Z}) avoids large parameter
changes, in regions poorly determined by the data, it contains very
little of our {\it ab initio} knowledge (and uncertainty) of how the functional 
should be.
   Therefore, the resulting functional, though optimal to fit a restricted set of
data, may be rather unplausible from a theoretical point of view,
and unreliable to reproduce other data.
   An obvious solution would be to minimize the error in the calculated 
energies, Eq.~(\ref{Z}), under a number of specific theoretical constraints, 
not just a general penalty term.
   Some of those constraints should be ``strict'', while others
may be more ``relaxed'' (quantitative).
   Thus, we know for sure that $F_x(0)=1$, but we are much less certain
about high values of $s$.
   More generally, the problem is to encode efficiently the known theoretical
information, either strict or ambiguous, and this is exactly the aim
of Bayesian probability theory.

   Bayes theorem can be succinctly stated as 
\begin{equation}
\PP(theory|facts) = C~\PP(facts|theory)\PP(theory), 
\label{Bayes}
\end{equation}
where $\PP(theory|facts)$
is the probability (or likelihood) that a theory is true, given that some facts 
have been observed, $\PP(facts|theory)$ is the probability that those facts 
would be observed if the theory was true, $\PP(theory)$ is (our estimate of) 
the {\it a priori} probability that the theory is true, and $C$ is a normalization
constant.
   In our case, ``theory'' means a quantitative parametrization of 
a given functional form (e.g. a set of GGA values $\epsilon_{\alpha\beta}$),
``facts'' are a set of $E^n_{ref}$ and $E^n_{GGA}$ energies, and ``true'' means
``optimal to reproduce the energies'' (not only of our dataset, but of
all possible systems of interest).

   Assuming a gaussian probability distribution,
\begin{equation}
\PP(facts|theory) = C_1 \exp \left( -\frac{1}{2} \sum_n  
          \left( \frac{E^n_{GGA}-E^n_{ref}}{\Delta E_n} \right)^2 \right)
\label{Pfacts_theory}
\end{equation}
where $\Delta E_n$ are the expected errors in the computed energies,
due to causes not related to the xc functional (e.g. basis set incompleteness), 
and to the inability of the GGA functional form to reproduce the exact energies.
   Equally
\begin{equation}
\PP(theory) = C_2 \exp \left( -\frac{1}{2} \sum_{\alpha\beta} 
                       (\epsilon_{\alpha}-\epsilon^0_{\alpha})
             C^{-1}_{\alpha\beta} (\epsilon_{\beta}-\epsilon^0_{\beta}) \right)
\label{Ptheory}
\end{equation}
where $\epsilon^0_\alpha$ is the average of $\epsilon_{xc}(k_{F\alpha},k_{G\alpha})$
among different {\it ab initio} GGA functionals:
\begin{equation}
\epsilon^0_\alpha = \sum_i w_i \epsilon^i_\alpha
\label{eps0}
\end{equation}
where index $i$ labels different functionals, 
$w_i$ are normalized weights assigned to them, and
$\epsilon^i_\alpha \equiv \epsilon^i_{xc}(k_{F\alpha},k_{G\alpha})$.
   $C^{-1}$ is the inverse of the covariance matrix
\begin{equation}
C_{\alpha\beta} = \sum_i w_i (\epsilon^i_\alpha-\epsilon^0_\alpha)
                             (\epsilon^i_\beta-\epsilon^0_\beta).
\label{Cov}
\end{equation}

\begin{figure}[!ht]
\includegraphics[width=0.8\columnwidth,clip]{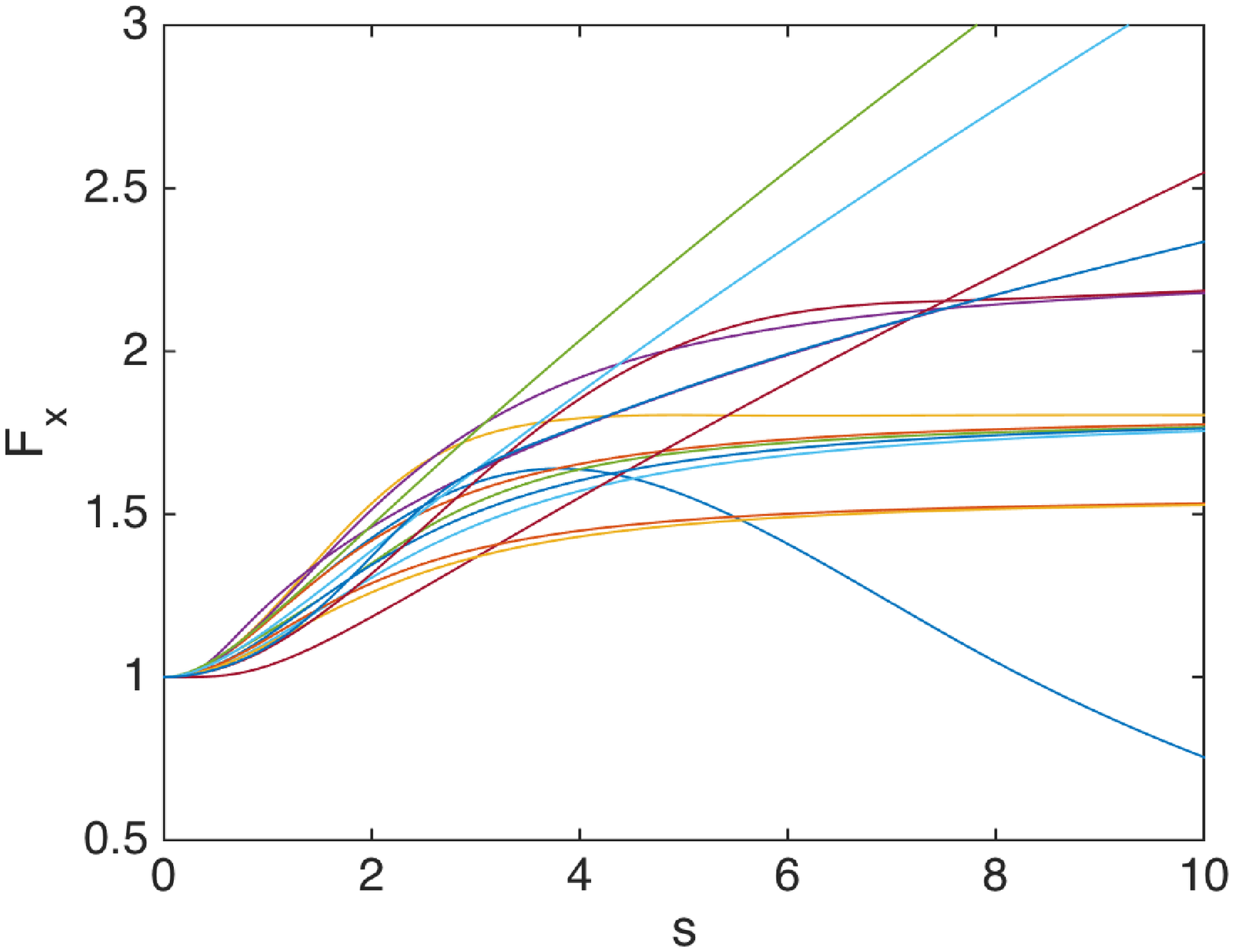}
\includegraphics[width=0.8\columnwidth,clip]{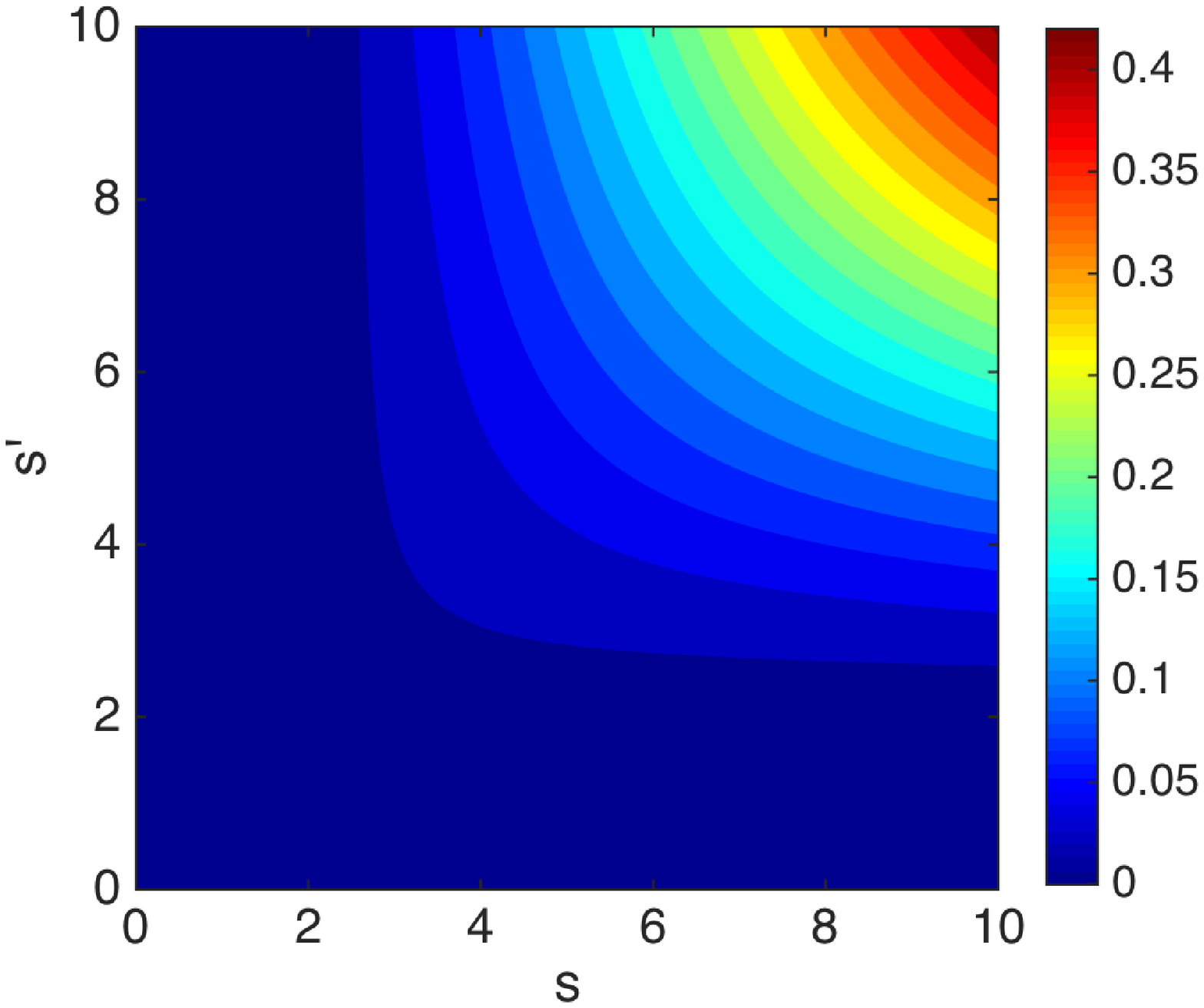}
\includegraphics[width=0.8\columnwidth,clip]{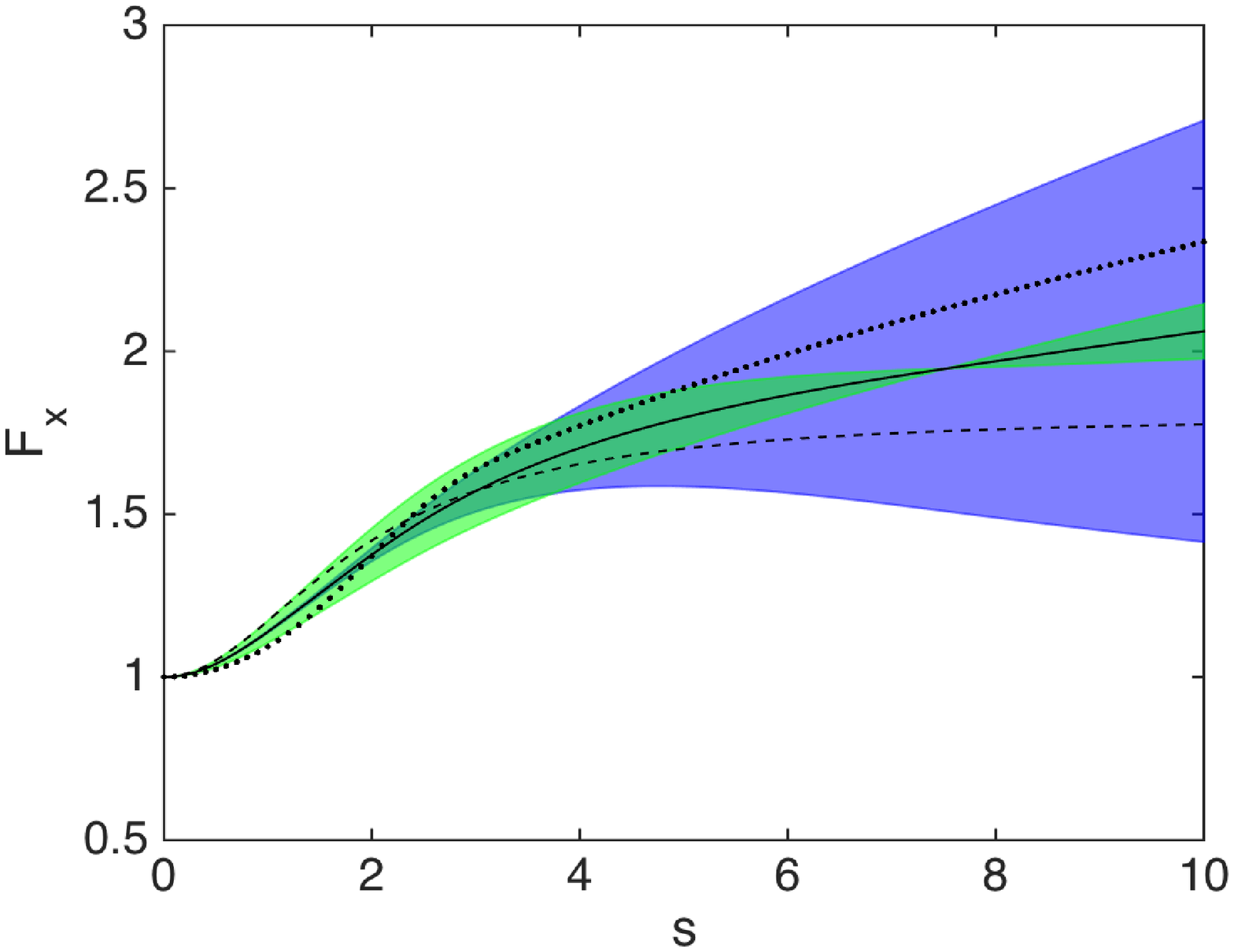}
\caption{
Top: {\it Ab initio} GGA exchange enhancement factors 
$F_x(s)$ used as prior information for the Bayesian optimization.
\cite{Perdew-Wang1986, Becke1988, Perdew-Burke-Ernzerhof1996, Zhang-Yang1998, 
      Hammer-Hansen-Norskov1999, Wu-Cohen2006, Perdew2008, Pedroza2009,
      Odashima2009, Mattsson-Armiento2009,
      Murray2009, Klimes2010, Cooper2010, Berland-Hyldgaard2014}
Middle: Contour plot of the covariance matrix $\text{Cov}[F_x(s),F_x(s')]$.
   Contour line values are multiples of 0.02.
Bottom: Average (continuous line) and first two eigenvectors of the covariance 
matrix (multiplied by the square root of their corresponding 
eigenvalues and shown as half the width of the blue and green areas). 
   Also shown are the enhancement factors of the 
PBE~\cite{Perdew-Burke-Ernzerhof1996} (dashed line) and
vdW-CX~\cite{Berland-Hyldgaard2014} (dotted line) functionals.
}
\label{sampleFx}
\end{figure}

   Notice that, in Eq.~(\ref{Ptheory}), we are using the discrepancies 
between the different ``{\it ab initio}'' functionals as a measure of 
our uncertainty on its exact form.
   However, the term $\PP(theory)$ in Eq.~(\ref{Bayes}) is important
mostly in regions of the functional domain that are poorly sampled
by the data, or in which the theoretical constraints are strict.
   In practice, the errors $\Delta E_n$ in Eq.~(\ref{Pfacts_theory})
can be used as a knob to balance our relative uncertainties on energy data,
on the one hand, and functional parameters, on the other.
   Notice also that the penalty approach of Eq.~(\ref{Z}) is equivalent to
using a diagonal covariance matrix for the functional parameters in the 
Bayesian approach.
   However, a diagonal covariance will not prevent the functional from 
developing strong oscillations, since the values of $\epsilon(k_F,k_G)$ 
at different points are assumed to be uncorrelated.
   In fact, the true covariance arising from a wide set of 
functionals is far from diagonal.
   Thus, Fig.~(\ref{sampleFx}) shows the average and the covariance of a
set of 15 parameterizations 
\cite{Perdew-Wang1986, Becke1988, Perdew-Burke-Ernzerhof1996, Zhang-Yang1998, 
      Hammer-Hansen-Norskov1999, Wu-Cohen2006, Perdew2008, Pedroza2009,
      Odashima2009, Mattsson-Armiento2009,
      Murray2009, Klimes2010, Cooper2010, Berland-Hyldgaard2014}
of the GGA exchange enhancement factor $F_x(s)$.
   It can be seen that the different functionals are highly correlated in
a rather complicated way, with a very non-diagonal covariance.

   Maximizing $\PP(theory|facts)$, i.e.
\begin{eqnarray}
Z &=& \sum_{n=1}^{N_{dat}} 
      \left( \frac{E^n_{GGA}(\eeps)-E_{refxc}^n}{\Delta E_n} \right)^2 \nonumber \\
  &+&  \sum_{\alpha\beta} 
           (\epsilon_{\alpha}-\epsilon^0_{\alpha})
           C^{-1}_{\alpha\beta} (\epsilon_{\beta}-\epsilon^0_{\beta}) = \min
\label{Zbayes}
\end{eqnarray}
with respect to the functional parameters $\epsilon_\alpha$ 
leads to a linear system of equations
\begin{equation}
\sum_{\beta} A_{\alpha \beta} ~\epsilon_\beta = B_\alpha
\label{bayesian_optimization}
\end{equation}
\begin{equation}
A_{\alpha \beta} = \sum_n \frac{\rho^n_\alpha \rho^n_\beta}{\Delta E_n^2}
                    + C^{-1}_{\alpha \beta}
\end{equation}
\begin{equation}
B_{\alpha} = \sum_n \frac{\rho^n_\alpha E_{refxc}^n}{\Delta E_n^2} +
             \sum_\beta C^{-1}_{\alpha \beta} \epsilon^0_\beta
\end{equation}
\begin{equation}
\rho^n_{\alpha} = \int d^3\rr ~\rho_n(\rr)
                  ~p_{F \alpha}(k_F(\rr)) ~p_{G \alpha}(k_G(\rr))
\end{equation}

   The resulting Bayesian method for functional optimization shares many
methodological characteristics of the BEEF method of 
Wellendorff {\it et al}~\cite{Wellendorff2012,Wellendorff2014,Medford2014},
as well as with that of Aldegunde {\it et al}~\cite{Aldegunde2016}.
   Like them, it has the ability to quantify the errors due to the
uncertainty of the resulting functional, although we will not discuss
this ability in the present work.
   However, the prior information in those methods is ``objective'', 
i.e. of general mathematical character, like a requirement
of smoothness for the resulting functional.
   In contrast, our method uses an ``informative'' prior, i.e. functional
constraints and uncertainties arising from a variety of {\it ab initio}
criteria.

\section{Data analysis in parameter space}

\begin{figure}[!ht]
\includegraphics[width=0.80\columnwidth,clip]{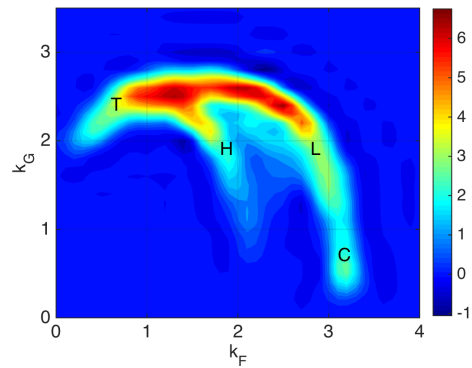}
\caption{
   Histogram of the electron density of the water molecule, in the
equilibrium geometry, as a function of wavevectors 
$k_F=(3\pi^2 \rho)^{1/3}$ and $k_G=|\nabla\rho|/\rho$. 
   The labels C, H, L, and T indicate regions of parameter space dominated
by the real-space regions of the oxygen core, hydrogen atoms, lone pairs, and 
electron tails, respectively.
   Atomic units are used.}
\label{rho_kF_kG_eq}
\end{figure}

\begin{figure}[!ht]
\includegraphics[width=0.80\columnwidth,clip]{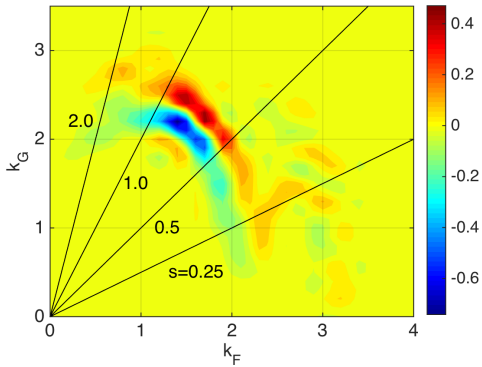}
\caption{
   Difference between the electron density $\rho(k_F,k_G)$ of the water 
molecule in the experimental geometry minus that in the relaxed geometry,
using the vdW-DF-cx functional in both cases.
   Also shown are lines of constant reduced gradient $s=k_G/2k_F$.
}
\label{delta_rho_kF_kG_monomer}
\end{figure}

   An important component of our approach is projecting the electron
density $\rho(\rr)$ of the data (i.e. of the water molecules) into 
``parameter space'' which, for a GGA functional, is that spanned by
wavevectors $k_F$ and $k_G$.
   For a given geometry, this is
\begin{eqnarray}
\rho(k_F,k_G) = \int d^3\rr  & \rho(\rr) &
   \delta \left( (3\pi^2 \rho(\rr))^{1/3}-k_F \right)  \nonumber \\
   & \times & \delta \left( \frac{|\nabla \rho(\rr)|}{\rho(\rr)}-k_G \right).
\label{rho_kF_kG}
\end{eqnarray}
   Figure~\ref{rho_kF_kG_eq} shows $\rho(k_F,k_G)$ for the isolated 
water molecule in its equilibrium geometry.
   It may be seen that the most relevant region of this parameter space is
that with $(k_F^2+k_G^2)^{1/2} \sim $~const. 
   We have observed that this is also true for a variety of systems, not
only water.
   However, what is generally most important are the {\em changes} of electron
density between different systems and geometries.
   Thus, Fig.~\ref{delta_rho_kF_kG_monomer} shows the change in 
$\rho(k_F,k_G)$ between the molecule in its equilibrium geometry 
with the vdW-DF-cx functional~\cite{Berland-Hyldgaard2014}
and with the experimental geometry.
   The basic difference between these two geometries is that, with vdW-DF-cx,
the OH bonds are $\sim 1.3$\% too long.
   This translates into a ``radial'' shift in $\rho(k_F,k_G)$, nearly
insensitive to the only variable $s$ of GGA exchange.
   This implies that optimizing $F_x(s)$ to reproduce the correct monomer 
geometry will force the functional to use rather artificial mechanisms, 
based on subtle changes of $\rho(k_F,k_G)$, to achieve the bond contraction.

\begin{figure}[!ht]
\includegraphics[width=0.80\columnwidth,clip]{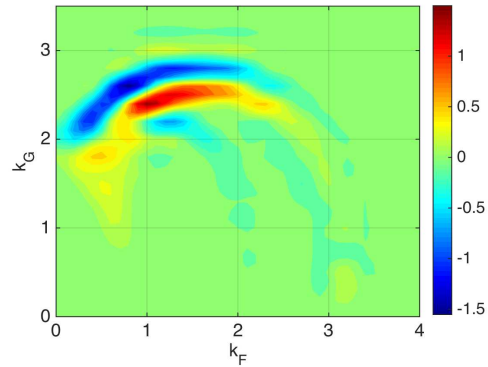}
\includegraphics[width=0.80\columnwidth,clip]{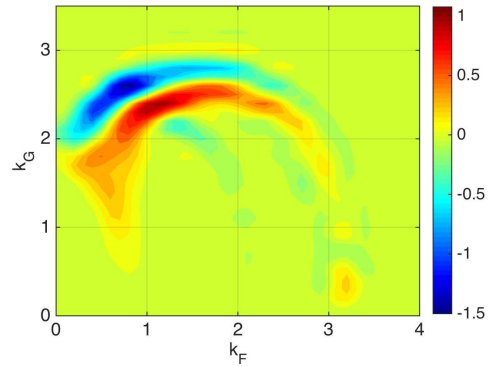}
\caption{
   Difference $\Delta \rho_{12} = \rho_{12}-\rho_1-\rho_2$ 
between the electron density $\rho(k_F,k_G)$ of a water 
dimer in its equilibrium geometry (top) and in a non-hydrogen-bonded geometry
(bottom), and that of the two separated monomers (in their dimer geometry).
}
\label{delta_rho_kF_kG_dimer}
\end{figure}

\begin{figure}[!ht]
\includegraphics[width=0.8\columnwidth,clip]{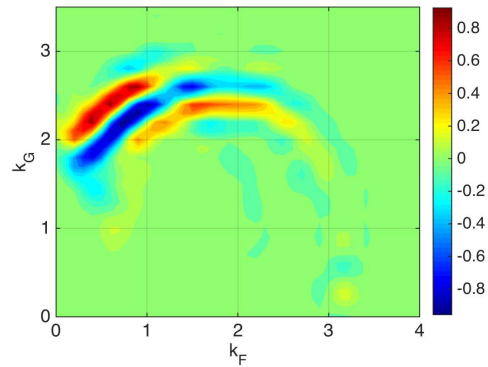}
\caption{
   Deformation density, defined by Eq.~(\ref{delta_rho_3b}), upon formation 
of a water trimer in its equilibrium geometry.
}
\label{delta_rho_kF_kG_trimer}
\end{figure}

    Fig.~\ref{delta_rho_kF_kG_dimer} shows the changes in 
$\rho(k_F,k_G)$ upon formation of a water dimer.
   In this case, the differences are more dependent on $s$,
and therefore more susceptible to the optimization of $F_x(s)$.
   Equally, Fig.~\ref{delta_rho_kF_kG_trimer} shows the deformation
density upon formation of a water trimer:
\begin{equation}
\Delta \rho_{123} = \rho_{123} - \rho_{12} - \rho_{23} - \rho_{13} 
    + \rho_1 + \rho_2 + \rho_3
\label{delta_rho_3b}
\end{equation}
where $\rho_i$, $\rho_{ij}$, and $\rho_{123}$ are the densities of the
monomers, dimers, and trimer, respectively.
   Comparison between Figs.~\ref{delta_rho_kF_kG_dimer} and 
\ref{delta_rho_kF_kG_trimer} shows that the dimer and trimer deformation 
densities are, to a large extent, opposite, implying that two- and three-body 
xc energies will have a large compensation.

\section{Optimization of a functional for water}

   Using the DPPS methodology, we have optimized a GGA exchange functional
for water.
   Of course, given the narrow scope of our system, we cannot assume that the 
resulting functional is significant for a broader range of DFT applications.
   Rather, our aim is to shed new light on why DFT has been so
frustratingly poor in simulating liquid water, on what 
changes are needed to improve it, and on the intrinsic shortcomings 
of the GGA form for this important system.

   For the DFT calculations, we have used a DPPS-enhanced version of the 
SIESTA code \cite{Soler2002}, with a quadruple-zeta plus double polarization
basis set~\cite{Corsetti2013a} and a highly converged 
mesh for real-space integrals and Fourier transforms.

   As prior information for the Bayesian optimization, we have used the 15
enhancement factors shown in Fig.~\ref{sampleFx}.
   For the reference energies, we use the MB-pol force 
field~\cite{Babin2013,Babin2014,Medders2014,Medders2015},
a sophisticated fit to highly accurate quantum chemical calculations
of water monomers and clusters.
   We selected geometries of 300 monomers, 330 dimers, and 100 trimers
for our training set.
   The monomers were generated with regular grids of bond
distances and angles.
   The dimers and trimers were taken from the MB-pol training set
(see SI for more information).
   It is important to emphasize that, although fitted only to monomers, dimers 
and trimers, the MB-pol force field has been shown to reproduce accurately the
structural and thermodynamic properties of condensed phases of 
water~\cite{Babin2013,Babin2014,Medders2014,Medders2015}.
   This shows that two- and three-body interactions are dominant in water,
and it gives us confidence that fitting (a large subset of) the MB-pol 
training set will also produce an accurate exchange functional for water.
   Details of these geometries and of our fitting weights and procedures 
can be found in the supporting information (SI)~\cite{SI}.

   Among the 15 functionals used as priors, some were designed for a limited
set of systems (e.g. bulk solids).
   Furthermore, while some are thoroughly tested and truly {\it ab initio},
others are more marginal, or they contain some degree of empirical information.
   Therefore, rather than the unweighted mean of the 15 functionals, 
we decided to use the newly proposed vdW-DF-cx\cite{Berland2014} 
as the reference functional ($\epsilon^0_\alpha$ parameters in 
Eqs.~(\ref{Z}) and (\ref{Ptheory})).
   This functional keeps the non local 
correlation form of the original vdW-DF\cite{Dion2004} and it proposes
a new consistent exchange (cx) which has shown to improve over the previous
versions of vdW-DF.

   Figure~\ref{fx_monomer} shows the exchange enhancement functions, 
optimized for water monomers, with and without Bayesian constraints
(Eqs.~(\ref{Zbayes}) and (\ref{Z}), respectively):
\begin{figure}[!ht]
\includegraphics[width=1.0\columnwidth,clip]{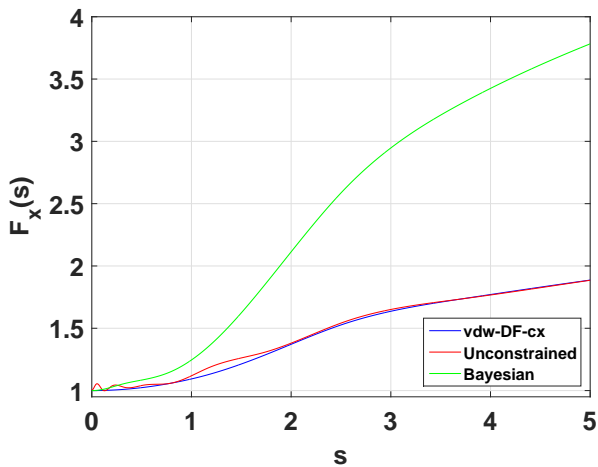}
\caption{
   GGA exchange enhancement factor $F_x(s)$, optimized to reproduce 
reference energies~\cite{Patridge-Schwenke1997} of the water monomer at
300 geometries (red).
   We show the result with (green) and without (red) bayesian constrains.
   The enhancement factor of the vdW-DF-cx functional (blue), used as the
reference functional in the optimization, is also shown for comparison.
}
\label{fx_monomer}
\end{figure}
   Also shown is the enhancement factor of the 
vdW-DF-cx functional~\cite{Berland-Hyldgaard2014}, 
used as reference for the initial functional parameters
(values $\epsilon^0_\alpha$ in Eqs.~(\ref{Z}) and (\ref{Ptheory})).
   Although the resulting energies are in excellent agreement with the 
reference data, the unconstrained functional shows strong oscillations,
that make it rather unplausible from a theoretical point of view.
   This is of course a consequence of not having imposed any physical
nor mathematical constraints on the plausible shapes of $F_x(s)$.
   When we do impose such constraints, by performing a Bayesian optimization
(Eq.~(\ref{Zbayes})) with {\it ab initio} functions $F_x(s)$ 
as prior information, the result is much more plausible, but there is a 
dramatic increase of $F_x(s)$.
   To understand this result, we notice in Fig.~\ref{delta_rho_kF_kG_monomer} 
that the correct density
$\rho(k_F,k_G)$ (corresponding to shorter OH distances) is shifted to
larger values of $k_G$ (and therefore, to larger $s$) for $s \sim 1$, 
in the region dominated by the electron tails.
   Therefore, a shorter bond is favoured by a larger slope of $F_x(s)$ 
for $s \sim 1$.
   And, since $F_x(s)$ is strongly constrained to be monotonous for 
$s<4$, the result is a large overall increase.

   The previous argument is in contrast with the usual result, that GGAs
increase binding distances relative to the LDA.
   This occurs because longer bonds decrease $k_F$ (the local
density) and increase $k_G$ (its gradient) in the bond region
(for an example, see the case of the oxygen molecule in SI).
   This increases $s$, what is favoured by the positive slope of $F_x(s)$ 
in the GGAs.
   The departure of H$_2$O from this usual behavior is due to the absence of 
a hydrogen core and to the small OH distance, with the hydrogen atom 
practically buried within the oxygen density.
   The main change, upon decreasing the OH distance,
is a global electronic contraction, with an average {\em increase} of the
density gradient around $s \sim 1$ (Fig.~\ref{delta_rho_kF_kG_monomer}).
   Therefore, the positive slope of $F_x(s)$ favors a shorter
OH distance, which {\em contracts} by 0.2\% with GGA-PBE, relative to the LDA.

\begin{figure}[!ht]
\includegraphics[width=0.80\columnwidth,clip]{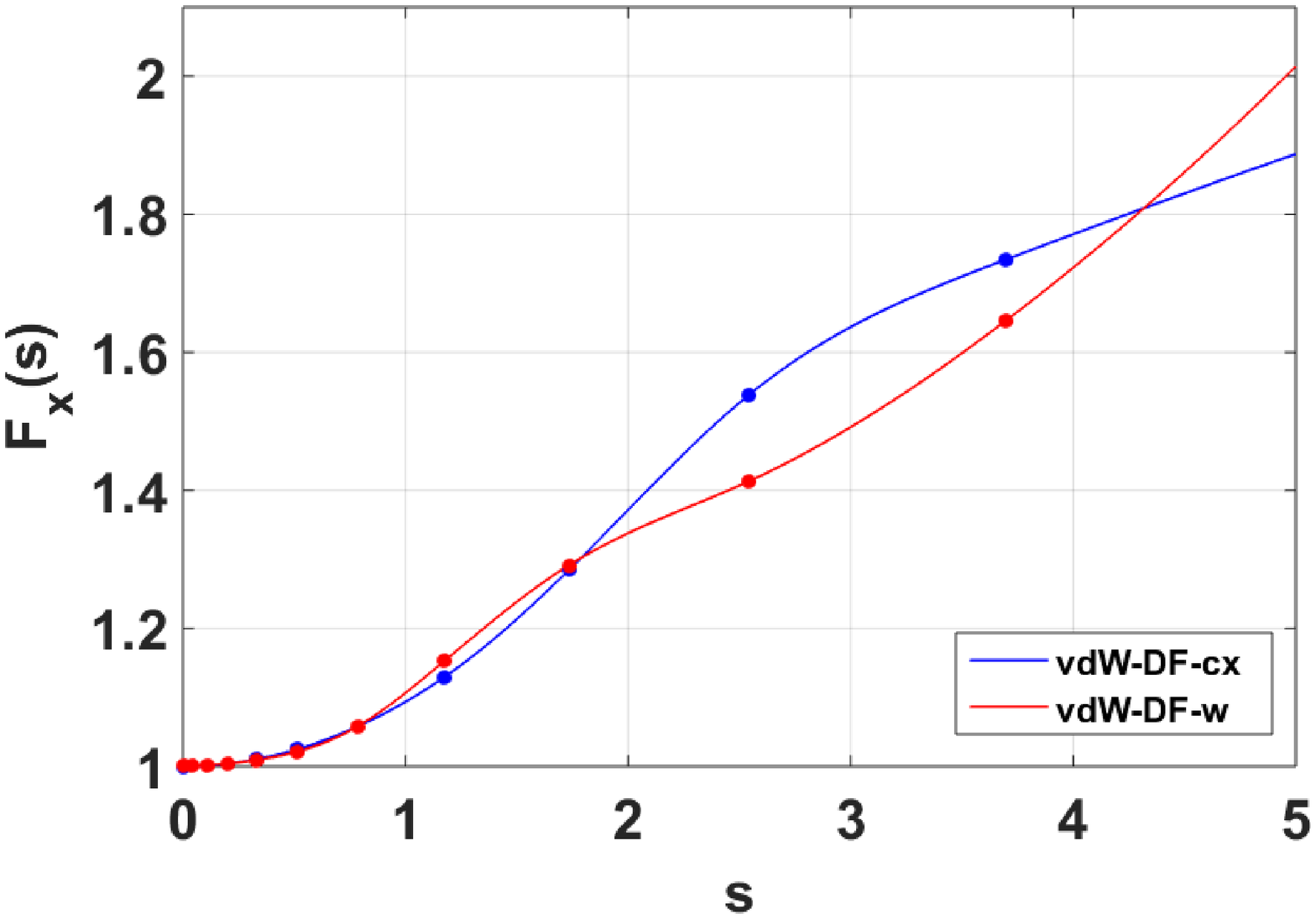}
\includegraphics[width=0.80\columnwidth,clip]{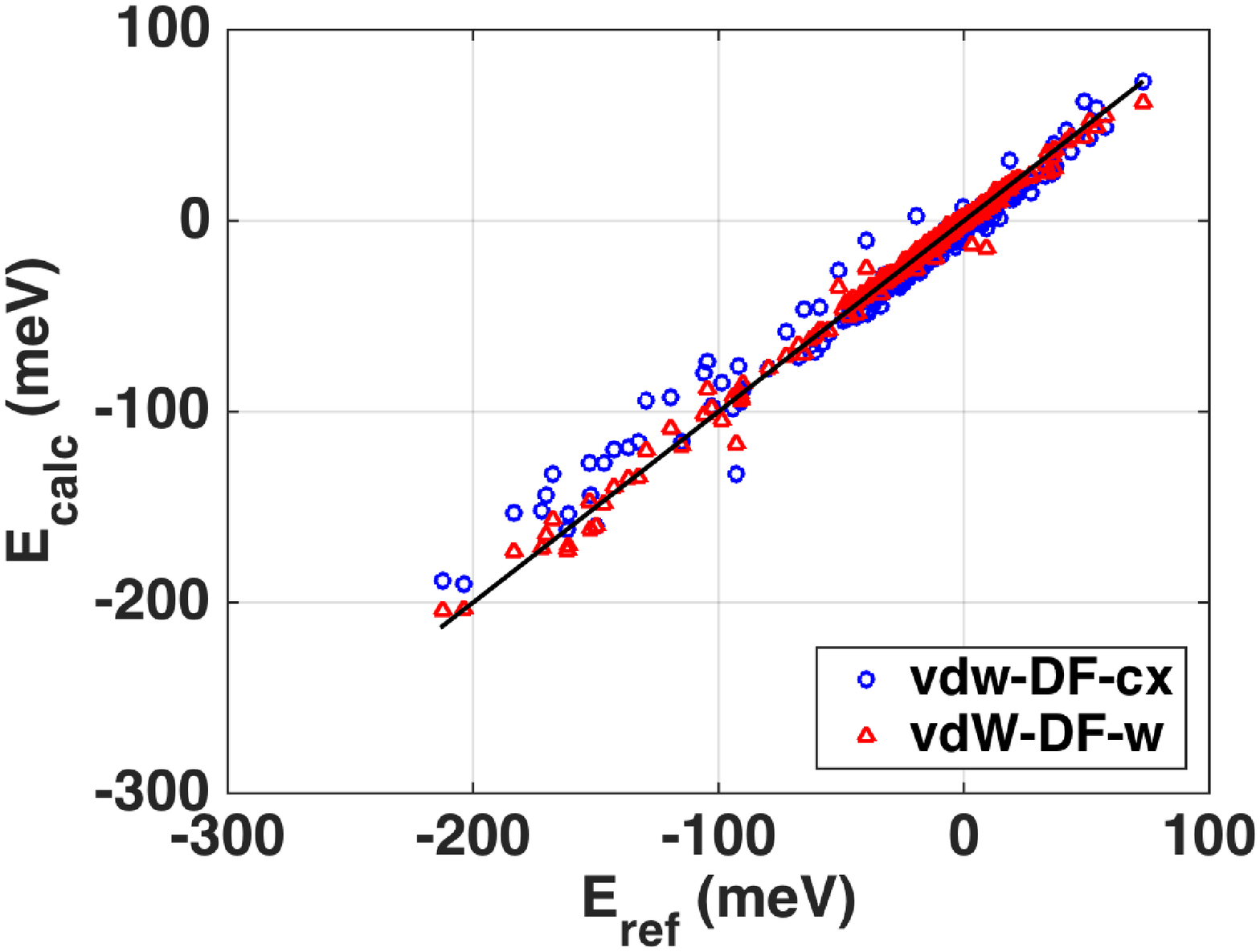}
\includegraphics[width=0.80\columnwidth,clip]{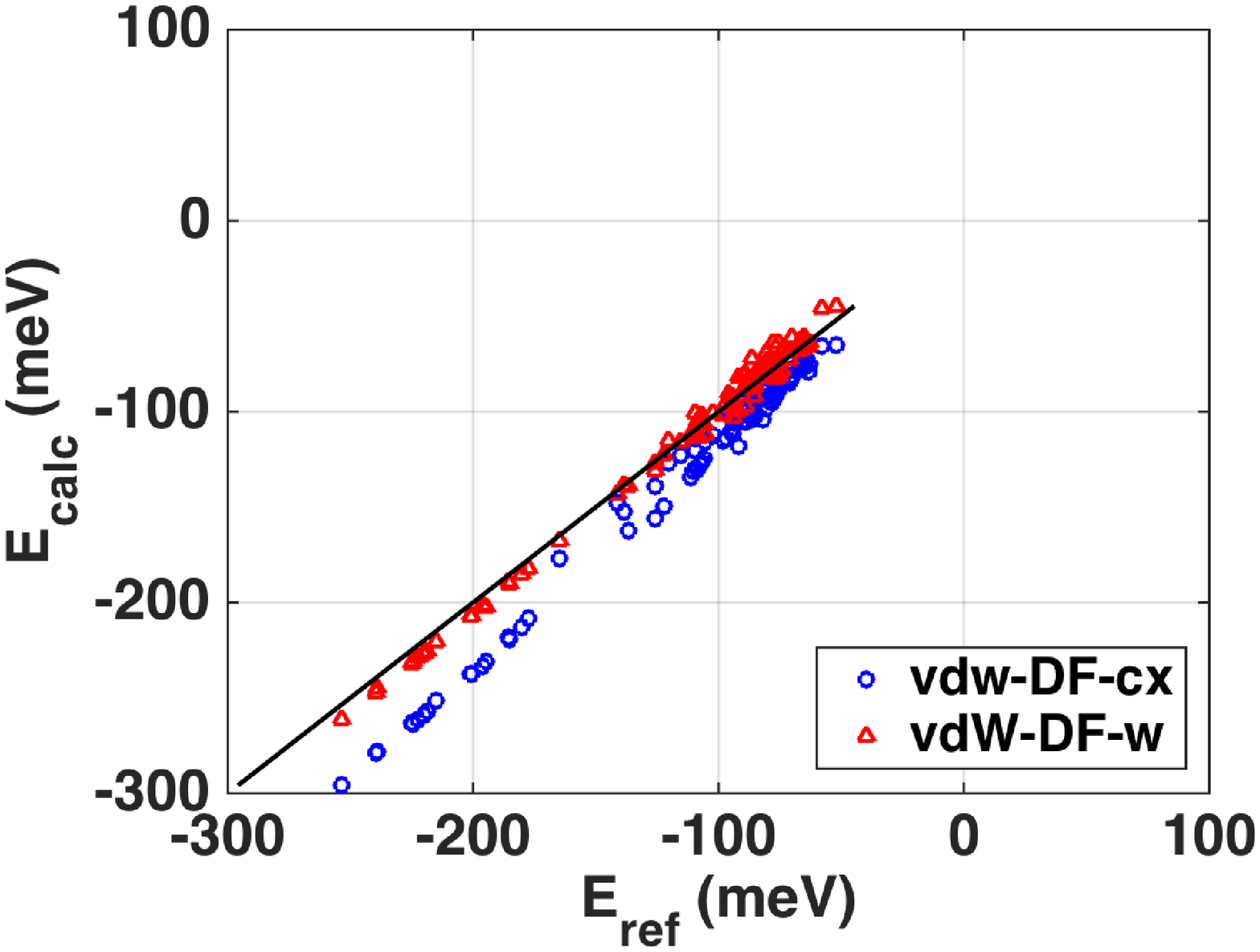}
\caption{
   Top: GGA exchange enhancement factor, optimized to reproduce 
reference energies of 330 water dimers and 100 trimers (red).
   Enhancement factor of the vdW-DF-cx functional (blue).
   Middle, red dots: comparison between the dimer interaction energies
$\Delta E_{12} = E_{12}-E_1-E_2$, calculated with the optimized functional, 
and the corresponding reference energies.
   The black line corresponds to a perfect agreement.
   The mean and mean square errors are
$\langle \Delta E \rangle = 0.02$ meV,
$\langle \Delta E^2 \rangle^{1/2} = 0.62$ meV.
   The blue dots are the same for the vdW-DF-cx functional, with
$\langle \Delta E \rangle = -0.17$ meV,
$\langle \Delta E^2 \rangle^{1/2} = 1.33$ meV.
   Bottom: the same for the trimer interaction energies
$\Delta E_{123} = E_{123}-E_{12}-E_{13}-E_{23}+E_1+E_2+E_3$.
   For the optimized functional,
$\langle \Delta E \rangle = 0.00$ meV,
$\langle \Delta E^2 \rangle^{1/2} = 0.60$ meV.
   For the vdW-DF-cx functional,
$\langle \Delta E \rangle = -1.85$ meV,
$\langle \Delta E^2 \rangle^{1/2} = 2.21$ meV.
}
\label{fx_custom_mb}
\end{figure}

   Although the Bayesian functional of Fig.~\ref{fx_monomer} is not itself 
physically unplausible, its shape is very different from
that required to fit the interaction energies of dimers and trimers
(Fig.~\ref{fx_custom_mb}).
   As a consequence, we have found essentially impossible to fit
simultaneously the monomer geometry and the interaction energies.
   This has led us to introduce an {\em ad hoc} correction for the 
energy difference between the exact and GGA energies of each monomer.
   This correction has the functional form of 
Ref.~\onlinecite{Patridge-Schwenke1997} and it adds a negligible overhead
to the DFT calculation.
   Similar monomer corrections have been previously applied by other 
authors\cite{Gillan2013, Bartok2013}. 
   Results assesing the qualty of this
1-body correction are presented in the SI\cite{SI}. 
   Although rather unsatisfactory from a theoretical point of view, this 
approach allows us to proceed with the optimization of the functional, in order 
to reproduce the more relevant energies of interaction between the different 
water molecules.
   In the future, we expect richer functional forms, like those of 
meta-GGAs, to correct this deficiency.

   Figure~\ref{fx_custom_mb} shows $F_x(s)$ optimized to reproduce the 
interaction energies of water dimers and trimers, using the Bayesian
method described above.
   The importance
of accurate 3-body term corrections to GGA functionals for water
has already been highlighted in Ref.~\onlinecite{Gillan2013}.
   The resulting functional also shows a dramatic improvement in the 
fitted energies, while the shape of $F_x(s)$ is still within the bounds 
of physical and mathematical plausibility.
   In the following, we will denote this functional vdW-DF-w, or
van der Waals density functional optimized for water.

   In addition to the energies of monomers, dimers, and trimers, 
the MB-pol force field was parameterized to reproduce also the electric
dipole $\mu$ and isotropic polarizability $\alpha$ of the molecule, 
which are important for the long range interactions.
   Therefore, it is important to notice that, although our vdW-DF-w 
functional was not trained to reproduce $\mu$ and $\alpha$,
it nevertheless does so rather well.
   Thus we find 
$\mu = 1.81$~D, $\alpha = $1.41~\AA$^3$ for vdW-DF-cx and
$\mu = 1.81$~D, $\alpha = $1.39~\AA$^3$ for vdW-DF-w, 
versus the reference values~\cite{Patridge-Schwenke1997,Babin2013} of 
$\mu =1.86$~D and $\alpha = $1.43~\AA$^3$.

\section{Results}

   In the following section we will present results for gas and condensed
phases of water, obtained with our optimized vdW-DF-w functional,
trained to reproduce 2- and 3-body energies from our training set. 
   For comparison, we will also present results for the vdW-DF-cx 
functional~\cite{Berland2014}, 
which has not yet been sufficiently evaluated for water.

\begin{figure}[!ht]
\includegraphics[width=0.80\columnwidth,clip]{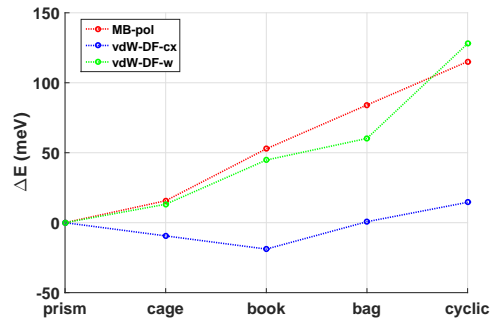}
\caption{
   Relative binding energies of five isomers of the H2O hexamer, whose geometries
are shown in the SI. 
   We show the energies for vdW-DF-cx, for our optimized
vdW-DF-w functional, and for the reference MB-pol force field.
}
\label{hexamer_energies}
\end{figure}

   As a first test of the ability of vdW-DF-w to describe
water systems beyond trimers, we show in Fig.~\ref{hexamer_energies} the
energies of various isomers of the water hexamer (whose geometries are
shown in SI).
   We can see that our functional does improve considerably the energies
of the vdW-DF-cx functional, relative to the MB-pol reference energies
(that reproduce accurately those of high-level quantum chemistry calculations).
   Thus, in contrast with vdW-DF-cx, our functional reproduces the 
correct order of the isomer binding energies.

   Because the small energy differences between its many solid structures,
the phase diagram of water ice is a major challenge for DFT.
   We have computed lattice energies for different ice polymorphs
at fixed geometries. 
   For the vdW-DF and MB-pol calculations, the geometries are those
given by Santra {\it et al}~\cite{Santra2013} for the PBE0-vdW-DF$^{TS}$
functional.
   For PBE, they are those given in the same reference for this functional.
   These geometries (given in SI) are rather different from those
of MB-pol, which results in an incorrect ordering of the different polymorphs.
   It is important to emphasize that, although no results have been reported
for relaxed MB-pol ice energies, a thorough comparison between MB-pol
and Quantum Monte Carlo energies for liquid snapshots demonstrated that
this force field yields much better agreement than any density functional
\cite{Medders2015}.
   In Table~\ref{IceTable} we compare our results with those of
MB-pol and of various other functionals. 
   It is significant that vdW-DF-w is the only functional that
reproduces qualitatively the behaviour of our reference MB-pol energies.
\begin{table} [ht!]
 \caption{
   Lattice energies of different ice phases, relative to ice Ih, in meV/molecule.
   All the calculations were done 
for fixed lattice and molecular geometries (provided in the SI).
   $\Delta_1$ indicates that the monomer correction described in the text
was included. 
}
\begin{center}
 		\begin{tabular}{l  c c c c c c c}
 		\hline
 		\hline
 		                & 	Ih 	&II	&IX	&XIII	&XIV	&XV	&VIII\\
 		\hline
 		vdW-DF-w			&0	&-11	&-14 &-22	&-23   &-7	 &11    \\
                vdW-DF-w-$\Delta_1$		&0	&-39	&-17 &-10	&-13   &-18	 &-5    \\
		vdW-DF-cx		        &0	&30	&14	 &20	&24	   &49   &90	\\
		vdW-DF-cx-$\Delta_1$		&0	&22	&11	 &31	&34    &38   &73	\\
		vdW-DF1         		&0	&12	& 8	 &13	&14    &23   &50	\\
		vdW-DF2             	&0	& 4	& 6	 & 5	& 5    &14   &33	\\
		PBE              		&0	&72	&51	 &83	&96    &114  &180	\\
		PBE\cite{Santra2013}   &0	&69	&49 	&80 	&93	&110	&177\\
 		MB-POL          		&0	&	&-37 &  	&-48   &-91  &-77	\\
		
 		\hline
 		\hline
 		\end{tabular}
 \end{center}
 \label{IceTable}       
 \end{table}

\begin{figure}[!ht]  
  \includegraphics[width=0.80\columnwidth,clip]{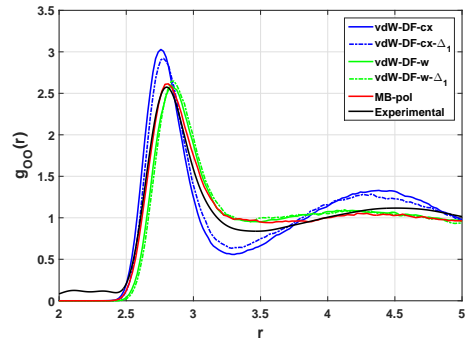}
  \caption{
     Oxygen-oxygen radial distribution function of water, at normal
  temperature-pressure conditions, with the vdW-DF-cx and optimized vdW-DF-w
  functionals, as well as with the reference MB-pol force field.
     Notice that our MB-pol simulation was performed under the same 
conditions as the DFT ones, and it is much shorter than that of 
Ref.~\onlinecite{Medders2014}.  
     The experimental RDF is also shown for comparison~\cite{Skinner2013}.
     Nuclear quantum effects were not included in any of the simulations.
  }
  \label{RDF}
\end{figure}

   It is well established that the competition between local structures,
with different average densities (generally called low- and high-density 
liquids, LDL and HDL), is the origin of abnormally high response functions 
and of most water anomalies\cite{Abascal2010,Xu2005}.
   This competition depends critically on a very subtle balance between the 
hydrogen-bond and van der Waals interactions, and it represents an extremely
difficult challenge for liquid water simulations in general, and for
{\it ab initio} molecular dynamics (AIMD) in particular.
   We have performed AIMD of 128 water molecules at normal temperature and
pressure conditions, using SIESTA with a double-zeta polarized basis set.
   Starting from the last geometry of a long Tip4P-2005~\cite{Abascal2005}
simulation, the system was thermalized with AIMD for 5 ps, and then sampled
for 10 ps.
   This procedure was followed for vdW-DF-cx and for our optimized vdw-DF-w 
functional, with and without correcting for the monomer energies.
   Nuclear quantum effects were not included.
   The resulting oxygen-oxygen radial distribution functions are compared in
Fig.~\ref{RDF} with experiment and with an 
MB-pol simulation without nuclear quantum effects.
   The later was performed under identical conditions and parameters 
as the AIMD simulations.
   It can be seen that our vdW-DF-w functional compares much better than
vdW-DF-cx with the reference MB-pol simulation.
   The optimized functional overcorrects the underestimation of O-O
distances (first peak of the RDF) by the vdW-DF-cx functional, giving
a right-shifted peak.

   The correction of the potential energy surface of the monomer dominates 
the total energies and it reduces the intramolecular OH distance by 1.3\%.
   From the well known anticorrelation between intra- and inter-molecular
bond distances and energies, it may be expected that the correction 
weakens the hydrogen bonds, and favors the HDL configurations.
   However, it can be seen that in practice the monomer correction has 
a rather small effect on the RDF.
   
\begin{figure}[!ht]
  \includegraphics[width=0.80\columnwidth,clip]{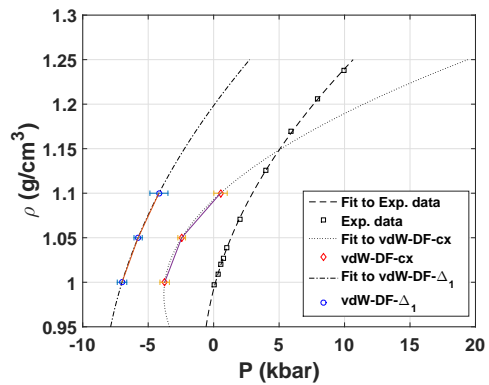}
  \caption{
     Comparison of the density-pressure curves from our liquid water 
simulations with vdW-DF-cx and vdW-DF-w and from experiment~\cite{Wagner2002}.
}
  \label{rho_vs_P}
\end{figure}

   Perhaps the most blatant shortcoming of DFT simulations of liquid water
is the huge range of discrepancies in the equilibrium density, ranging from
$\sim 0.7$ g/cm$^3$ for some GGAs to $\sim 1.15$ g/cm$^3$ for some vdW-DFs
\cite{Wang2011}.
   The pressure-density curves from our simulations 
are compared with experiment in Fig.~\ref{rho_vs_P}.
   It may be seen that vdW-DF-cx and vdW-DF-w overestimate the equilibrium
density by 10\% and 20\%, respectively.
   In the latter, however, this is because of a rigid shift of the curve
to lower pressures, so that the compressibility (shown in SI) is in good 
agreement with experiment.
   Since the pressure is also very sensitive to finite size and nuclear
quantum effects (not included in our simulations), as well as to
basis set superposition errors (the DZP basis used in the simulations is
smaller than that used in the functional optimization), it remains to
be studied to what extent the density overestimation is due to these effects.

   Another important magnitude, that depends strongly on the ratio of 
LDL to HDL in the actual liquid, is the self diffusion constant.
   The experimental value of 0.23~\AA$^2$/ps is strongly underestimated 
by GGAs under normal conditions of density and temperature~\cite{Wang2011}.
   For the vdW-DF-cx functional, we find 0.08~\AA$^2$/ps. 
   When we include the monomer energy correction, this increases slightly 
to 0.09~\AA$^2$/ps, still much lower than the experimental value.
   With vdW-DF-w, we obtain 0.17~\AA$^2$/ps without the monomer correction
and 0.23~\AA$^2$/ps with it, in agreement with experiment.

\section{Conclusions}

   In conclusion, we have designed and implemented a general Bayesian method 
to optimize an exchange-correlation functional.
   It combines {\it a priori} information from {\it ab initio} functional
constraints, with a database of reference energies and geometries.
   Using this method, we have optimized for water a GGA exchange functional,
combined with the correlation functional of vdW-DF~\cite{Dion2004}.
   As prior information, we have used the covariance of 15 GGA exchange 
functionals.
   As reference, we used a database of hundreds of monomers, dimers, 
and trimers taken from the MB-pol training 
set~\cite{Babin2013,Babin2014,Medders2014,Medders2015}.
   We find that the optimization of the monomer geometry is largely 
incompatible with that of intermolecular interactions.
   As an {\it ad hoc} solution, we have developed a correction for the
potential energy surface of the monomer, and we have optimized the
functional to reproduce the dimer and trimer interaction energies.
   The resulting functional performs considerably better than previous
GGA and vdW-DF functionals for extended clusters, ices, and liquid water.

   We acknowledge Prof. M. Gillan for discussions and for providing an
initial database of MP2 energies and geometries.
   We also acknowledge Prof. F. Paesani for the MB-pol code and for
assistance in its installation and use.
   This work has been funded by MINECO grants FIS2012-37549 and FIS2015-64886
and by DOE grants DE-FG02-09ER16052 and DE-SC0003871.


\begin{thebibliography}{49}
\bibitem[{spl()}]{splines}
\bibinfo{note}{In the cubic splines method, it is also necessary to specify the
  two boundary conditions (either a given value of the first derivative,
  usually zero, or a zero second derivative}.
\end{thebibliography}

\end{document}